\documentstyle[prl,aps,twocolumn,epsfig]{revtex}

\begin{document}

\wideabs{
\title{Trapped-Ion Quantum Simulator: Experimental Application to Nonlinear Interferometers}

\author{D. Leibfried, B. DeMarco,  V. Meyer,  M.
 Rowe$^*$, A. Ben-Kish$^{\dag}$, J. Britton, W. M. Itano,\\ B. Jelenkovi\'{c}$^\sharp$,
 C. Langer, T. Rosenband and D. J. Wineland}

\address{Time and Frequency Division, National Institute of Standards and Technology, Boulder, Colorado 80305}
\date{\today}
\maketitle

\begin{abstract}
We show how an experimentally realized set of operations on a
single trapped ion is sufficient to simulate a wide class of
Hamiltonians of a spin-1/2 particle in an external potential. This
system is also able to simulate other physical dynamics. As a
demonstration, we simulate the action of an $n$-th order nonlinear
optical beamsplitter. Two of these beamsplitters can be used to
construct an interferometer sensitive to phase shifts in one of
the interferometer beam paths. The sensitivity in determining
these phase shifts increases linearly with $n$, and the simulation
demonstrates that the use of nonlinear beamsplitters ($n$=2,3)
enhances this sensitivity compared to the standard quantum limit
imposed by a linear beamsplitter ($n$=1).
\end{abstract}

\pacs{03.67.-a, 03.67.Lx, 42.50.-p} }

One of the motivations behind Feynman's proposal for a quantum
computer \cite{Feynman85} was the possibility that one quantum
system could efficiently simulate the behavior of other quantum
systems. This idea was verified by Lloyd \cite{Lloyd96} and
further explored by Lloyd and Braunstein \cite{Lloyd99} for a
conjugate pair of variables such as position and momentum of a
quantum particle. Following this suggestion we show below that
coherent manipulation of the quantized motional and internal
states of a single trapped ion using laser pulses can simulate the
more general quantum dynamics of a single spin-1/2 particle in an
arbitrary external potential. Previously, harmonic and anharmonic
oscillators have been simulated in NMR \cite{Somaroo99}.

In addition to demonstrating the basic building blocks for
simulating such arbitrary dynamics, we experimentally simulated
the action of optical Mach-Zehnder interferometers with linear and
nonlinear second- and third-order beam-splitters on number-states.
Interferometers with linear beamsplitters and nonclassical input
states have engendered considerable interest, since their noise
limits for phase estimation can lie below the standard quantum
limit for linear interferometers with coherent input modes
\cite{Caves80,Caves81,Yurke86,Holland93} as has been demonstrated
in experiments \cite{Xiao87}. A number of optics experiments have
exploited the second-order process of spontaneous parametric
down-conversion \cite{Tittel01}, which can be regarded as a
nonlinear beamsplitter.  By cascading this process, a fourth-order
interaction has also recently been realized \cite{Lamas01}. One
difficulty in these experiments is the exponential decrease in
efficiency as the order increases, necessitating data
post-selection and long integration times.  In the simulations
reported here, nonlinear interactions were implemented with high
efficiency, eliminating the need for data post-selection and
thereby requiring relatively short integration times.

To realize a quantum computer for simulating a spin $s=1/2$
particle of mass $\mu$ in an arbitrary potential, one must be able
to prepare an arbitrary input state
\begin{equation}\label{ArbSta}
  |\Psi(m_s,z)\rangle=
  \sum_{n} (c_{\downarrow n} | \downarrow \rangle|n \rangle+
  c_{\uparrow n} | \uparrow \rangle|n \rangle),
\end{equation}
where the particle's position wavefunction is expanded in energy
eigenstates $|n \rangle$ of a suitable harmonic oscillator and
$|m_s \rangle$ ($m_s \in \{ \downarrow, \uparrow \}$) represent
the spin eigenstates in a suitable basis. We have recently
demonstrated a method to generate arbitrary states of the type in
Eq.(\ref{ArbSta}) in an ion trap \cite{Law96,BenKish02}. The
computer should then evolve the state according to an arbitrary
Hamiltonian
\begin{eqnarray}\label{ArbHam}
  H&=& \left[\frac{p^2}{2 \mu}+ V(z,m_s)\right] \nonumber \\ &\simeq&
   \sum_{n,m\leq n}^{N} (\alpha_{nm}I+\beta_{nm}\sigma_+ + \beta^*_{nm}\sigma_-
  +\gamma_{nm}\sigma_z)\nonumber \\
  &\times& (\chi_{nm} (a^\dag)^n a^m + \chi^*_{nm}(a^\dag)^m a^n),
\end{eqnarray}
where we require only that the potential $V(z,m_s)$ can be
expanded as a power series in the harmonic oscillator ladder
operators $a$ and $a^\dag$ and be approximated to arbitrary
precision by a finite number of terms with maximum order $N$. The
$m_s$ are a set of observables in a general two-level Hilbert
space that can all be mapped to a linear combination of the
indentity $I$ and the Pauli matrices $\sigma_j$. The operators
$\sigma_\pm$ are defined as $\sigma_\pm =\sigma_x \pm i \sigma_y$,
all $\beta_{nm}, \chi_{nm}$ are complex numbers, and  all
$\alpha_{nm}, \gamma_{nm}$ are real numbers.

In our realization of an analog quantum computer we consider the
Hamiltonian of a trapped atom of mass $\mu$, harmonically bound
with a trap frequency $\omega_z$ and interacting with two
running-wave light fields having a frequency difference $\Delta
\omega$ and a phase difference $\varphi$ at the position of the
ion. Both light fields are assumed to be detuned from an excited
electronic level so they can induce stimulated-Raman transitions
between combinations of two long-lived internal electronic
ground-state levels with energy difference $\hbar \omega_0$ and
the external motional levels of the ion\cite{Wineland98a}. For our
purpose it is sufficient to consider the motion along one axis in
the trap. After applying a rotating wave approximation and
adiabatic elimination of the near resonant excited state
\cite{Wineland98a}, and switching to an interaction picture of the
ion's motion, the resonant interaction for Raman beam detuning
$\Delta \omega =\epsilon \hspace{1pt} \omega_0 + l \omega_z$
($\epsilon=\{0,1\}$, $l$ integer) can be written in the Lamb-Dicke
limit ($\eta^2 \langle (a+a^\dag)^2 \rangle \ll 1$) as
\cite{Wineland98b,Matos98}
\begin{equation}\label{ResHam}
    H_{\epsilon l}=\hbar \Omega e^{i \phi}(\sigma_+)^\epsilon
\left[\delta_{l,|l|} \frac{(i \eta a)^{|l|}}{|l|!}
+(1-\delta_{l,|l|}) \frac{(i \eta a^\dag)^{|l|}}{|l|!} \right] +
h.c.
\end{equation}
The coupling strength $\Omega$ is assumed to be small enough to
resonantly excite only the $l$-th spectral component. The
Lamb-Dicke parameter $\eta= \Delta k~z_0$ is the product of the
$z$-projection of the wavevector-difference $\Delta k$ of the two
light fields and the spatial extent of the ground state wave
function $z_0=\sqrt{\hbar/(2 m \omega_z)}$. For $\epsilon=1$ the
internal state changes during the stimulated Raman transition and
the interaction couples $| \downarrow \rangle|n \rangle
\leftrightarrow | \uparrow \rangle | n+ l\rangle$, while for
$\epsilon=0$ only motional states $|n \rangle \leftrightarrow |n+l
\rangle$ are coupled with a strength independent of the internal
state \cite{Footnote1}.

Following Lloyd and Braunstein \cite{Lloyd99,Footnote2}, by
nesting and concatenating sequences of $H_{\epsilon l}$ operations
according to the relation
\begin{equation}\label{ConCat}
  e^{-\frac{i}{\hbar}H \delta t} e^{-\frac{i}{\hbar}H' \delta
  t} e^{\frac{i}{\hbar}H \delta t} e^{\frac{i}{\hbar}H' \delta
  t}= e^{\frac{1}{\hbar^2}[H,H'] \delta t^2}+O(\delta t^3),
\end{equation}
the set of operators
$\{H_{01},H_{02},H_{03},H_{10},H_{11},H_{12},H_{13}\}$ is
sufficient to efficiently generate arbitrary Hamiltonians. This
conclusion is straightforward for the spin, since $\{\sigma_+,
\sigma_-,\sigma_z\}$ are a complete basis of that algebra. For
interactions that only involve the motion ($\epsilon=0$) it
follows from the fact that
\begin{equation}\label{ComRel}
  \left[H_{02},H_{03}\right] \propto
  i \left\{\alpha a^\dag a^2 + \alpha^* (a^\dag)^2 a  \right\} +{\rm lower~orders}
\end{equation}
and
\begin{eqnarray}\label{ComRe2}
 \left[\alpha a^\dag a^2 + \alpha^* (a^\dag)^2 a,\beta (a^\dag)^n a^m + \beta^* (a^\dag)^m
 a^n \right] = \nonumber \\
 (2m-n)\left[\alpha \beta (a^\dag)^{m}a^{n+1}+\alpha \beta^*
 (a^\dag)^{n}a^{m+1}-h.c.\right]
 \nonumber \\
 +{\rm ~lower~orders},
\end{eqnarray}
so one can build up arbitrary orders in the effective Hamiltonian
by recursive use of Eq.(\ref{ConCat}). Similar arguments hold for
the set of $\{H_{1l}\}$ interactions, and by combining both types
of interactions, the series expansion of the Hamiltonian in
Eq.(\ref{ArbHam}) can eventually be constructed.

Most of these interactions have been demonstrated in previous
ion-trap experiments. $H_{10}$ is usually called the carrier
interaction, $H_{01}$ and $H_{02}$ are coherent and squeeze drives
respectively and $H_{11},H_{12}$ are first and second blue
sideband \cite{Meekhof96,Leibfried97}. The third-order
interactions $H_{03},H_{13}$ have not been previously
demonstrated. One of the experiments discussed below uses two
$H_{13}$ pulses, therefore demonstrating the feasibility of
generating $H_{03}$ as well \cite{Footnote3}.

\begin{figure}
\begin{center}
\epsfig{file=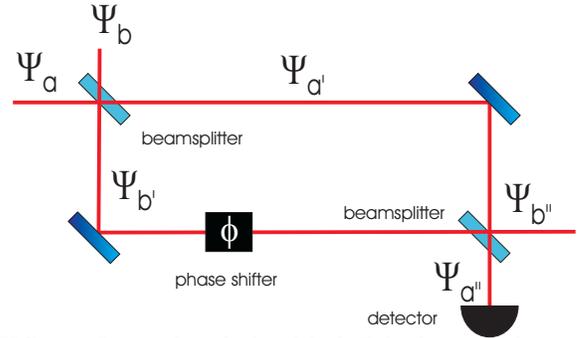, width=7.5 cm, clip}
\caption{\label{fig:MacZeh} Principle of the Mach-Zehnder
interferometer. Modes $\Psi_a$ and $\Psi_b$ are superposed on the
first beamsplitter. After the beamsplitter has acted the modes
$\Psi_{a'}$, $\Psi_{b'}$ are propagated along separate paths to a
second beamsplitter. Mode $\Psi_{b'}$ may undergo a variable phase
shift induced by the phase shifter $\phi$. Modes $\Psi_{a''}$ and
$\Psi_{b''}$ emerge after the second beamsplitter and one of the
modes is put onto a detector. Varying $\phi$ will lead to a
sinusoidal behavior of the intensity on the detector (fringes).}
\end{center}
\end{figure}

As a demonstration of quantum simulation using a single trapped
atom, we employ the interactions $H_{11}, H_{12}$, and $H_{13}$ to
efficiently simulate a certain class of $n$-th order optical
beamsplitters described by Hamiltonians
\begin{equation}\label{BeaSpl}
B_n=\hbar \Omega_n [a(b^\dag)^n + a^\dag(b)^n].
\end{equation}
Here $a$ and $b$ are the usual harmonic oscillator lowering
operators for the two quantized light modes, $\Omega_n$ is the
coupling strength, and we simulate the special case where the
number of photons in mode $a$ is 0 or 1 and $n=1,2$ or 3. Two such
beamsplitters can be used to construct a Mach-Zehnder
interferometer as sketched in Fig. \ref{fig:MacZeh}. The order
$n=1$ corresponds to the commonly used linear beamsplitter that is
typically realized by a partially transparent mirror in
experiments. Such interferometers can measure the relative phase
of the two paths of the light fields that are split on the first
beamsplitter and recombined on the second. The phase can be varied
by changing a phase shifting element (the box labeled $\phi$ in
Fig. \ref{fig:MacZeh}) and detected (modulo $2 \pi$) by observing
the interference fringes of the recombined fields. We restrict our
attention to a pure number-state $|n=1 \rangle_a$ impinging on the
first beamsplitter from mode $a$ and a vacuum state $|n=0
\rangle_b$ from mode $b$. After propagating the input state
through the first beamsplitter with $\Omega_n$ adjusted to give
equal amplitude along the two paths in the output superposition,
the state becomes
\begin{equation}\label{IntPr1}
    |1\rangle_a|0 \rangle_b  \rightarrow \frac{1}{\sqrt{2}}(|1 \rangle_{a'} |0 \rangle_{b'} + |0 \rangle_{a'} |n
  \rangle_{b'}).
  \end{equation}
Phase shifters in optical interferometers alter a classical-like
coherent state $|\alpha\rangle$ to one that is shifted to $|\alpha
e^{i \phi} \rangle$. In the context of Fig.1 this phase shift
corresponds to $|n \rangle \rightarrow e^{i n \phi} |n \rangle$
for a number state, leading to
\begin{eqnarray}\label{IntPr2}
  \frac{1}{\sqrt{2}}(|1 \rangle_{a'} |0 \rangle_{b'} + |0 \rangle_{a'} |n
  \rangle_{b'})
  \rightarrow \nonumber \\
  \frac{1}{\sqrt{2}}(|1 \rangle_{a'} |0 \rangle_{b'} + e^{i n \phi} |0 \rangle_{a'} |n
  \rangle_{b'}).
\end{eqnarray}
The second beamsplitter recombines the two field modes leading to
an average probability of
\begin{equation}\label{IntInt}
  \langle \hat{n}_{a''}\rangle=\frac{1}{2}[1-\cos(n \phi)]
\end{equation}
for detecting one photon in the output arm with the detector in
Fig.\ref{fig:MacZeh}.

We have experimentally simulated the nonlinear beam-splitter of
Eq.(\ref{BeaSpl}) using a single trapped $^9$Be$^+$ ion. The
operator $a$ is replaced by $\sigma^+$, the raising operator
between two hyperfine states $|F=2,m_F=-2\rangle
\equiv|1\rangle_a$ and $|F=1,m_F=-1\rangle \equiv|0\rangle_a$ in
the $^2$S$_{1/2}$ ground state manifold. These operators (and also
their respective Hermitian conjugates) are not strictly
equivalent, but their action is the same as long as we restrict
our attention to situations that never leave the $\{|0
\rangle_a,|1 \rangle_a \}$ subspace. The simulated linear and
nonlinear interferometers fulfill this restriction, as long as the
input state  is $|1\rangle_a|0 \rangle_b$. The optical mode with
lowering operator $b$ is replaced by the equivalent harmonic
oscillator mode of motion along one axis in the trap, with number
states $|n\rangle$.

\begin{figure}
\begin{center}
\epsfig{file=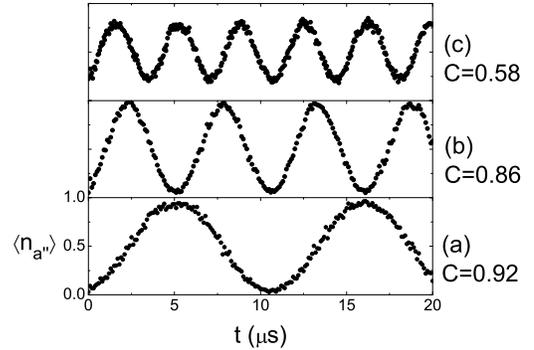, width=7.5 cm, clip}
\caption{\label{fig:IntFri} Interference fringes for simulated
interferometers (a) of order $n=1$, (integration time per data
point was 0.50 s); (b) $n=2$, (0.53 s) and (c) $n=3$, (0.63 s).
$\langle n_{a''}\rangle$ is the probability to find the ion in
$|1\rangle_{a''}$, while $t$ is the time for which the trap
frequency was shifted by $\Delta \omega_z$, directly proportional
to the phase shift $\phi=\Delta \omega_z~t$. The frequency of the
fringes increases linearly with order $n$. $C$ is the observed
contrast of the fringes.}
\end{center}
\end{figure}

Our experimental system has been described in detail elsewhere
\cite{Meekhof96,Leibfried97,Sackett01}. We trapped a single
$^9$Be$^+$ ion in a linear trap \cite{Rowe02} with motional
frequency $\omega_z= 2 \pi $ 3.63~MHz (Lamb-Dicke parameter
$\eta=$0.35) and cooled it to the ground state of motion. The trap
had a heating rate of 1 quantum per 6 ms \cite{Rowe02} that was a
small perturbation for the duration of our experiments ($\leq$ 260
$\mu$s). After cooling, the ion was prepared in the $|1\rangle_a|0
\rangle_b$ state by optical pumping. Starting from this state we
used Raman-transitions to drive a $\pi/2$-pulse on the ion's
$n$-th blue sideband ($H_I \propto \sigma^+ (b^\dag)^n+h.c.$),
creating the state
$(|1\rangle_{a'}|0\rangle_{b'}+|0\rangle_{a'}|n\rangle_{b'})/\sqrt{2}$.
For different orders $n$ the $\pi/2$-pulse time scales as
$\sqrt{n!}/\eta^n$ \cite{Wineland98a}. The observed $\pi/2$-times
of $(4.0,17.3,115)~\mu$s do not scale exactly as the theoretical
prediction due to different laser intensities used for the
different values of $n$. A phase shift $\phi=\Delta \omega_z~t$
was then introduced by switching the potential of the trap endcaps
to a different value for time $t$, thus changing the motional
frequency by a fixed amount $\Delta\omega_z$. After a second
$\pi/2$-pulse on the $n$-th sideband we measured the probability
$\langle n_{a''}\rangle$ for the ion to be in $|1\rangle_a$. The
interference fringes created by sweeping $t$ are shown in Fig.
\ref{fig:IntFri}. The final state of the ion oscillated
approximately between $|1\rangle_{a''}$ and $|0\rangle_{a''}$ as
$t$ was varied, with frequency $n \Delta\omega_z$.

\begin{figure}
\begin{center}
\epsfig{file=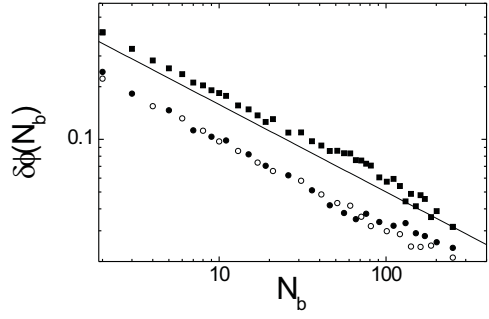, width=7.5 cm, clip}
\caption{\label{fig:AllVar} Noise-to-signal ratios $\delta \phi
(N_b)$ for the $n=1$ linear interferometer (solid squares), $n=2$
nonlinear interferometer (solid circles) and $n=3$ nonlinear
interferometer (open circles) vs. the number of measurements
$N_b$. The solid line is the theoretical limit for the linear
(n=1) interferometer, assuming perfect contrast and detection
efficiency.}
\end{center}
\end{figure}

In interferometric measurements, we want to maximize our
sensitivity to changes of $\phi$ around some nominal value. We
therefore want to minimize
\begin{equation}\label{DelPhi}
  \delta\phi=\frac{\Delta \hat{n}_{a''}}{|\partial\langle
  \hat{n}_{a''}\rangle/\partial\phi|},
\end{equation}
where $\Delta \hat{A}\equiv\sqrt{\langle \hat{A}^2\rangle- \langle
\hat{A}\rangle^2}$ is a measure of the fluctuations between
measurements of an operator $\hat{A}$. Eq. (\ref{DelPhi}) applies
to our simulator with $\phi=\Delta\omega_z t$. In our experiments
\begin{equation}\label{DowExp}
\langle \hat{n}_{a''}\rangle=(C/2)~[1-\cos(n \phi)],
\end{equation}
where $C$ is the contrast of the observed fringes. Ideally $C=1$
[Eq. (\ref{IntInt})] but is observed to be $<1$ due to imperfect
state preparation and detection, and fluctuations in the ambient
magnetic field and the trap frequency . The sensitivity of the
interferometer is maximized when the slope of $\langle
\hat{n}_{a''} \rangle$ with respect to $\phi$, $\partial\langle
\hat{n}_{a''}\rangle/\partial \phi$ is maximized, that is, for
values of $\phi$ where $n \phi= \pi k/2$, $k$ an odd integer. We
characterize the fluctuations $\Delta n_{a''}$ with the two-sample
Allan variance, commonly used to characterize frequency stability
\cite{Allan66}. In the present context, a series of $M$ (total)
measurements of $\hat{n}_{a''}\equiv|1\rangle_{a''}\langle
1|_{a''}$ is divided into bins of $N_b$ measurements averaged
according to
\begin{equation}\label{AllMea}
\langle\hat{n}_{a''}\rangle_i=1/N_b
  \sum_{j=iN_b}^{(i+1)N_b-1}(\hat{n}_{a''})_j,
\end{equation}
where $2<N_b<M/2$ and $(\hat{n}_{a''})_j$ is the $j$-th
measurement of $\hat{n}_{a''}$. The Allan variance characterizing
fluctuations between measurements is given by
\begin{equation}\label{AllVar}
  (\sigma_{\hat{n}_{a''}}(N_b))^2\equiv
  \frac{1}{2(N_b-1)}\sum_{i=1}^{N_b-1}
  (\langle\hat{n}_{a''}\rangle_{i+1}-\langle\hat{n}_{a''}\rangle_{i})^2.
\end{equation}
Making the identification $\sigma_{\hat{n}_{a''}}(N_b)=\Delta
n_{a''}$, in Fig. \ref{fig:AllVar}, we plot $\delta \phi$ vs.
$N_b$. The solid curve is the theoretical standard quantum limit
for a linear interferometer with perfect contrast and unity
detection efficiency, given by $\Delta n_{a''}/\sqrt{N_b}$ where
$(\Delta n_{a''})^2$ is the variance due to projection noise
\cite{Itano93}; $\Delta n_{a''}=0.5$ at the points of maximum
slope in our fringes. The simulation of the linear interferometer
shows only a small amount of excess noise over the theoretical
limit, due mainly to the $C=0.92$ contrast of the fringes, while
the nonlinear interferometer simulations have a noise-to-signal
ratio below the linear interferometer standard limit. The
potential gain in slope for $n=3$ is almost exactly cancelled by
the loss in fringe contrast, so the noise-to-signal ratio for
$n=2$ and $n=3$ is about the same.

In conclusion, we have shown how coherent stimulated-Raman
transitions on a single trapped atom can be used to simulate a
wide class of Hamiltonians of a spin-1/2 particle in an arbitrary
external potential.  This system can also be used to simulate
other physical dynamics. As a demonstration, we have
experimentally simulated the behavior of $n$-th order nonlinear
optical beam-splitters acting in a restricted Hilbert space. Our
simulation demonstrates how interferometer sensitivity improves
with the order of the beam splitter.  As a practical matter, the
2nd- and 3rd-order beamsplitters demonstrated here give increased
sensitivity for diagnosing motional frequency fluctuations in the
trapped-ion system.  With anticipated improvements in motional
state coherence \cite{Rowe02}, it should be possible to simulate
more complicated Hamiltonians.

The authors thank M. Barrett and D. Lucas for suggestions and
comments on the manuscript. This work was supported by the U. S.
National Security Agency (NSA) and Advanced Research and
Development Activity (ARDA) under contract No. MOD-7171.00, the
U.S. Office of Naval Research (ONR). This paper is a contribution
of the National Institute of Standards and Technology and is not
subject to U.S. copyright.\\
\\
\noindent $\dag$ current address: Department of Physics, Technion, Haifa, Israel.\\
\noindent $*$ current address: Optoelectronics Division, NIST, Boulder, CO, USA.\\
\noindent $\sharp$ Institute of Physics, Belgrade, Yugoslavia.\\



\begin{thebibliography}{99}
\bibitem{Footnote1}
The coupling can be experimentally arranged to depend or not
depend on the internal state, see \cite{Wineland98a}.
\bibitem{Footnote2}
Note that Eq.(1) in \cite{Lloyd99} has a sign error in the
right-hand side exponent.
\bibitem{Footnote3}
$H_{13}$ and $H_{03}$ both scale as $\eta^3 \Omega$, and differ
only by $\omega_0$ in the Raman-detuning. Therefore sucessful
impementation of $H_{13}$ implies the feasibility of $H_{03}$.
\end{thebibliography}
\end{document}